\documentstyle[version2,aps]{revtex}
\begin{document}
\begin{title}
\center{Collective modes of spin, density, phase and amplitude in
        exotic superconductors}
\end{title}
\author{D.van der Marel}
\begin{instit}
Laboratory of Solid State Physics, Department of Science\\
University of Groningen, Nijenborgh 4, 9747 AG Groningen
\end{instit}
\begin{abstract}
The equations of motion of pair-like excitations in the superconducting
state are studied for various types of pairing using the random
phase approximation. The collective modes are computed of a layered
electron gas described by
a $t-t'$ tight-binding band, where the electrons experience besides
the long-range Coulomb repulsion an on-site
Hubbard U repulsion and a nearest-neighbour attractive interaction.
{}From numerical calculations we see, that the collective
mode spectrum now becomes particularly rich. Several
branches can occur below the continuum of quasi-particle excitations,
corresponding to order-parameter fluctuations of various symmetries
of pairing, and collective spin-density fluctuations.
{}From the collective mode softening near the nesting vectors it
is concluded, that in the d-wave paired state an instability occurs toward the
formation of a spin-density wave.
\vspace{2\baselineskip}\\
Materials Science Center Internal Report Number VSGD.94.6.7
\vspace{1\baselineskip}
\end{abstract}

\widetext

\section{Introduction}
A well-known result of BCS theory is the
variational wavefunction, describing the
ground state of a superconductor. In the limit
$\vec{Q}\rightarrow 0$ this function can be easily extended to
describe a superconductor\cite{bcs,tinkham} moving at
a small and uniform velocity $\vec{v}=(2m_e)^{-1}\hbar \vec{Q}$
\begin{equation}
|\Psi> =
\left[\int d^3\vec{R} e^{i\vec{Q}\cdot\vec{R}} \int d^3\vec{r}
 \phi(\vec{r})\psi^{\dagger}_{\uparrow}(\vec{R}+\vec{r}/2)
\psi^{\dagger}_{\downarrow}(\vec{R}-\vec{r}/2)\right]^{N/2}|0>
\label{eq:psi}
\end{equation}
This function has the mathematical shape of a Bose condensate of pairs, where
the wavefunction $\phi(\vec{r})$ describing the relative motion
of electrons forming a pair is the Fourier transform of
$\left\{[1+(\epsilon_k/\Delta_k)^2]^{-1/2}-(\epsilon_k/\Delta_k)\right\}$,
and $\exp{(i\vec{Q}\cdot\vec{R})}$
is the macroscopic wavefunction describing the center of mass motion
of each pair.
The similarity to a Bose condensate wavefunction is somewhat misleading,
as also the wavefunction of a gas of uncorrelated
fermions can be written in this form,
in which case $\phi(\vec{r})$ is a non-trivial
function with an $r^{-2}$ tail.
In the limit of a weak effective interaction $\phi(\vec{r})$
has an algebraic tail just as for the free electron gas. If
the interaction is strong,
$\phi(\vec{r})$ can be interpreted as a wave-function describing the
relative motion of two electrons forming a Bose-condensed pair.\cite{nsr}
If the effective interaction
is an on-site attraction, the electrons pair-up in a singlet-wave
function with an enhanced probability to occupy the same site.
Clearly if the electrons experience a strong on-site repulsion,
the tendency towards pairing disappears. With a net attraction between
electrons occupying neighbouring sites in the lattice, it is still
possible to form a paired state, but $\phi(r)$ has to be constructed such,
that the particles avoid the same site. This condition is for example
fulfilled when when $\phi(\vec{r})$ has a finite angular momentum.
\\
One may wonder whether the analogy to Bose-condensation can be drawn
further, and consider the energy spectrum of pair-like excitations as
a function of pair-momentum. This problem was first
treated by Bogoliubov\cite{bogoliubov}, and Anderson\cite{pwa58}.
If the electrons experience an on-site
repulsion, with a nearest-neighbour attraction, the collective
mode spectrum becomes particularly rich. It turns out that several
branches occur below the continuum of quasi-particle excitations,
corresponding to order-parameter fluctuations of various symmetries
of pairing\cite{bardasis}. The existance
of low-lying collective modes may be important when attempting to
identify a superconducting gap in the
infrared, Raman, or inelastic neutron scattering spectra of these materials.
\\
Collective modes in superconductors have in the past attracted the attention
for a variety of reasons:
(1) Bogoliubov predicted the existance of a longitudinal collective mode
with a sound-like dispersion\cite{bogoliubov}. Long
range Coulomb interactions make the spectrum identical to the plasmons of
a normal Fermi gas, as was shown by Anderson\cite{pwa58}.
(2) The collective mode spectrum naturally follows from a gauge invariant
formulation of BCS theory \cite{pwa58}, and a consistent
explanation of the Meissner effect requires that the whole interaction
Hamiltonian (as opposed to the reduced BCS Hamilitonian) is
taken into account\cite{pwa58,rickayzen}.
(3) As collective modes mediate electron-electron interactions,
plasmons\cite{morawitz,bose,cote} and spin-fluctuations
\cite{monthoux93,scalapino,levin,pao} have been considered as possible
candidates
for a pairing-mechanism.
(4) Certain modes, in particular condensate phase-fluctuations near or
below the
pair-breaking gap, are important for the thermal behaviour,
notably $T_c$, of the superconductor\cite{kirchberg1}.
(5) An instability of the ground state and an incipient phase transition
to a state with a lower energy follow from the softening of collective
modes\cite{vaks,kostyrko,micnas}.
(6) Collective modes may show up in experimental spectra, such as in
optical\cite{hirashima,hirschfeld,doniach} or Raman spectroscopy
\cite{varma,monien}.
(7) As there is no inter-plane hopping in a layered electron gas,
the $k$-dependent plasmon spectrum becomes gapless\cite{fetter}, which
may give rise to an interesting behaviour in the region for momentum and
frequency values where the collective mode crosses $2\Delta$\cite{fertig,cote}.
(8) If there exists an electron-electron interaction in channels with
a finite angular momentum $L$, excitons with the corresponding
symmetries can exist\cite{bardasis,vaks,hirashima}.\\
Usually modes of phase and density\cite{fertig,cote}
are treated separately from amplitude modes\cite{varma,monien}, and
spin-fluctuations\cite{scalapino,monthoux93,spinsuper}.
As we will see below, especially for a non-vanishing momentum
a coupling exists between the four collective-mode channels of
spin-density, charge, phase and amplitude of the order parameter. The
aim of this study is, to derive general expressions for the collective
modes in the superconducting state, using a unified approach
including effects of finite momentum pairing. In the last section
examples are given for the collective modes and the generalized
susceptibility in the superconducting state. It is shown that a
$d$-wave superconductor may become unstable with respect to the
formation of a spin-density wave, or possibly a mixed SDW-wave plus
superconducting state, if an on-site repulsion is taken into account
in addition to having an attractive interaction in the $d$-wave channel.
More detailed calculations of various response functions
and the comparison thereof to measurements on specific
materials will follow in a future publication.
\section{The model hamiltonian}
In the discussion of the collective modes we will make extensive use
of two-particle creation operators. We will see below, that
the channels with $S_z=-1$, $S_z=1$ and $S_z=0$ are decoupled. In the
$S_z=\pm 1$ channels there are triplet pair-excitations, and
spin-fluctuations. In the $S_z=0$ channel there are spin-fluctuations, density
fluctuations, and fluctuations of phase and amplitude of the
order-parameter (singlet and triplet pair-excitations).
The corresponding operators are in the same order
\begin{equation}
\begin{array}{lll}
\sigma_k(Q) &\equiv& c_{k+Q \uparrow}^{\dagger} c_{k \uparrow}
           - c_{-k+Q\downarrow}^{\dagger} c_{-k\downarrow} \\
\rho_k(Q)   &\equiv& c_{k+Q \uparrow}^{\dagger} c_{k \uparrow}
           + c_{-k+Q\downarrow}^{\dagger} c_{-k\downarrow} \\
\phi_k(Q)   &\equiv& c_{-k-Q \downarrow}c_{k \uparrow}
           - c_{k+Q \uparrow}^{\dagger} c_{-k\downarrow}^{\dagger} \\
\psi_k(Q)   &\equiv& c_{-k-Q \downarrow}c_{k \uparrow}
           + c_{k+Q \uparrow}^{\dagger} c_{-k\downarrow}^{\dagger} \\
\end{array}
\end{equation}

The remaining 6 combinations are
$c_{k+Q \sigma}^{\dagger} c_{k \underline{\sigma}}$,
(spin-fluctutuations with $S_z=\pm 1$),
and $c_{-k-Q \sigma}c_{k \sigma}$ with the corresponding hermitian conjugates
(spin-triplet phase- and amplitude-fluctuations).
When transformed to Euclidean-space representation these operators
acquire a more transparant physical meaning. For example the
spin density distribution function $n_{\uparrow}(r)-n_{\downarrow}(r)$
has as its Fourier transform $\sum_k\sigma_k(Q)$.
Similar relations exist for the other operators, and the
notation $\sigma(Q)$, $\rho(Q)$, $\phi(Q)$, and $\psi(Q)$, will
be used to indicate the Fourier-transforms of the spin-density,
charge density, phase and gap-amplitude distribution in Euclidean
space.
We  will consider a system of interacting electrons which can be
described with the following Hamiltonian:
\begin{equation}
H = \sum_{k} \xi_k\rho_k(0) +
   \sum_{Q}\left\{\frac{1}{2} V(Q) \rho(Q) \rho(-Q)
                  +\frac{1}{8} U(Q) \left[ \rho(Q) \rho(-Q)
                  -\vec{\sigma}(Q) \cdot \vec{\sigma}(-Q) \right]
                   \right\}
\label{eq:hamiltonian}
\end{equation}
where in $V(Q)$ I lumped together the Coulomb interaction with all other
spin-independent interactions, which could be due to the coupling of the
electrons to the other degrees of freedom of the solid. In principle, and
in particular if the interaction Kernel is derived from boson-exchange
models such as electron-phonon interaction, there can also be
a separate dependence on the momentum of the interacting particles.
For compactness of notation
I will not explicitly include such a $k$ and $q$ dependence in the
Hamiltonian. \\
With the spin-dependent
interaction assumed here, the total spin of the system is still a
good quantum number.
Such terms can appear if the model Hamiltonian is derived from a more
fundamental one by projecting out part of the Hilbert space. A
well-known example is the occurance of the Kondo exchange-interaction
in a magnetic impurity system after carrying out the Schrieffer-Wolff
transformation. Other examples where such terms occur are the RKKY
interaction in magnetic alloys, and the superexchange in rare earth-doped
semiconductors. Also the on-site Hubbard U term is usually written
in this form, although in this case the Pauli-principle already
automatically excludes occupation of the same site with parallel spins.
\\
As we will discuss the equations of motion of the collective modes for
a general form of the effective electron-electron interaction, it is
worthwhile to summarize the expressions for the gap equation and the
free energy. The thermodynamic potential at $T=0$ of a BCS superconductor is
the expectation value of the grand canonical Hamiltonian, and is
easily obtained by taking the expectation value of Eq. \ref{eq:hamiltonian}
using the variational wave function of Eq.\ref{eq:psi}
\begin{equation}
\Omega(\mu,V,v_{k_1},...,v_{k_N}) =
    2\sum_k|v_k|^2 \left(\xi_k-\mu \right)
   +  \sum_{kq} \left(u_k v_k \lambda_{kq} u^*_q v^*_q
   + |u_k|^2 V(k-q)|v_q|^2 \right)
\label{eq:rickomega}
\end{equation}
where $\lambda_{kq}$ is the pairing potential. For the
type of interaction introduced above one obtains
$\lambda_{kq}=V(k-q)+\frac{1}{2}U(k-q)+\frac{1}{2}U(k+q)$ where the
last term is the spin-flip scattering contribution contained
in $\vec{\sigma}_k(Q) \cdot \vec{\sigma}_q(-Q)$. The last
term in Eq. \ref{eq:rickomega} corresponds to the exchange
energy. From $\Omega$
one obtains the gap-equation by calculating the minimum as a function of
the set of variational parameters ${v_{k_1},...,v_{k_N}}$. The
number of particles in the ground state is obtained
by taking the first derivative with respect to $\mu$. The resulting set
of equations is
\begin{equation}
\begin{array}{lll}
-\sum_q u_q v_q \lambda_{kq}&=&\frac{2u_kv_k}
{|u_k|^2-|v_k|^2}\tilde{\epsilon}_k\\
              \sum_k |v_k|^2&=&N_e                                          \\
\end{array}
\label{eq:gap1}
\end{equation}
Apart from a shift in chemical potential the effect of the exchange energy term
on the thermodynamic potential is to renormalize the single particle dispersion
$\epsilon_k=\xi_k-\mu$, which now has to be replaced with
$\tilde{\epsilon}_k\equiv\epsilon_k-\sum_q|v_q|^2V(k-q)$.
After the ground state has been found from minimalization of the
free energy, the quasi-particle spectrum is obtained, with the usual BCS-type
energy dispersion $E_k=\left[\tilde{\epsilon}^2+\Delta_k^2\right]^{1/2}$, and
with $\Delta_k/E_k$ defined as $2u_kv_k$. In the
following sections I will also use the (standard) notations $b_k\equiv u_kv_k$
and $z_k\equiv (|u_k|^2-|v_k|^2)/2$. \\
If $\lambda_{kq}$ has a non-trivial $k$-dependence we can make
a partial-wave decomposition
\begin{displaymath}
\lambda_{kq}=\sum_{\alpha}\psi^*_{\alpha}(k)
\lambda_{\alpha}\psi_{\alpha}(q)
\end{displaymath}
where $\{\psi_{\alpha}(k)\}$ is a complete set of orthogonal
functions, chosen such as to diagonalize the pairing
potential. We can make a similar expansion of the order parameter
$\Delta_k=\sum_k\Delta_{\alpha} \psi_{\alpha}(k)$ with the
help of which one obtains the coupled gap equations
\begin{displaymath}
\Delta_{\alpha}=-\lambda_{\alpha}
 \sum_k \sum_{\beta} \frac{\psi^*_{\alpha}(k)
 \psi_{\beta}(k)\Delta_{\beta}}{2E_k}
\end{displaymath}
We notice, that for $\Delta \rightarrow 0$ a decoupling of pairing-channels
occurs, depending on the presence of off-diagonal elements in
the decomposition of $1/E_k \rightarrow 1/|\epsilon_k|$.
As $\lambda_{kq}$ is real,
the set $\{\psi_{\alpha}(k)\}$ can be chosen as real numbers. As a result
also $\Delta_{\alpha}$ is real. Solutions like "s+id"\cite{kotliar} become
possible if there is a degeneracy between solutions with a different
symmetry.
\section{Equations of motion}
The equations of motion are of the form
$[H,\hat{O}]=\nu \hat{O}$, where $\hat{O}$ is a linear combination
of pair-operators representing an
excitation of the system with energy $\nu$.
Although these equations have been treated extensively before, in the
previous papers the coupling to the collective spin oscillation
channel has not been considered. In particular a spin-dependent
term was not included in earlier publications. As one of the aims of
this paper is to discuss collective modes of spin-density in the
superconducting state,
I re-derive the equations of motion with this extended hamiltonian. \\
In the superconducting state the equations of motion of
spin density($\sigma_k(Q)$), charge density ($\rho_k(Q)$), order parameter
phase ($\phi_k(Q)$), and order parameter amplitude ($\psi_k(Q)$)
are coupled in a non-trivial way.
The commutator of each of these two-particle operators with the
interaction part of the Hamiltonian generates
products of four single-particle operators, which are approximated by taking
the expectation value of all combinations of two of the operators appearing
in this product. The resulting terms fall in two categories: those which
have the same $k$-value, and those which are a weighted summation over
$k$-space. The latter give rise to the collective modes. In the first category
one obtains (1) self energy
terms which can be absorbed in a shift of the chemical potential,
(2) exchange self energy terms, due to which $\epsilon_k$ is renormalized to
$\tilde{\epsilon}_k \equiv \epsilon_k - \sum_q |v_q|^2 V(k-q)$, and
(3) cross-terms proportional to $\Delta_k$, linking $\sigma_k$
to $\psi_k$, and $\rho_k$ to $\phi_k$-operators.
\\
Finally the category of weighted averages of two-particle operators
over $k$-space
involves both direct and exchange terms, and is
given by the expressions
\begin{equation}
\begin{array}{lll}
{\bf S}_k(Q)  &\equiv& \sum_q  H^i_{\sigma}(k,q,Q) \sigma_q(Q) \\
{\bf R}_k(Q)  &\equiv& \sum_q  H^i_{\rho}(k,q,Q) \rho_q(Q)  \\
{\bf A}_k(Q)  &\equiv& \sum_q  H^i_{\phi}(k,q,Q) \phi_q(Q)  \\
{\bf B}_k(Q)  &\equiv& \sum_q  H^i_{\psi}(k,q,Q) \psi_q(Q)  \\
\end{array}
\label{eq:srab}
\end{equation}
where I introduced
\begin{equation}
\begin{array}{lll}
H^i_{\sigma}(k,q,Q) &\equiv& - \frac{1}{2}U(Q) - V(k-q) - \frac{1}{2}U(k-q) \\
H^i_{\rho}(k,q,Q)   &\equiv& 2V(Q)
                             + \frac{1}{2}U(Q) - V(k-q) + \frac{1}{2}U(k-q) \\
H^i_{\phi}(k,q,Q)   &\equiv& V(k-q) + \frac{1}{2}U(k-q) + \frac{1}{2}U(k+q) \\
H^i_{\psi}(k,q,Q)   &\equiv& V(k-q) + \frac{1}{2}U(k-q) + \frac{1}{2}U(k+q) \\
\end{array}
\label{eq:hi1}
\end{equation}
With these definitions, and using the random phase approximation
described above, the commutators of the pair operators can now
be derived. The actual calculation is a straightforward, though rather
laborious, exercise in commutator algebra. A detailed description of
the various terms has been given by Anderson, and later discussed
more extensively by Bardasis and Schrieffer, who retained a number of
vertices in their final analysis which were neglected by Anderson. In
the present paper all vertices discussed by Bardasis and Schrieffer
are taken into account. The expressions are
however modified due to the spin-dependent interaction term in
Eq. \ref{eq:hamiltonian}.
The set of commutators, including the exchange interactions, is
\begin{equation}
\begin{array}{llll}
{[H,\sigma_{k}(Q)]} &=&
   \tilde{\epsilon}^-_{kQ}\rho_{k}(Q)
 - \Delta_{kQ}^-\psi_k(Q)
 &       + z^-_{kQ} {\bf R}_k(Q)
         - b^-_{kQ} {\bf B}_k(Q)   \\
{[H,\rho_{k}(Q)]} &=&
   \tilde{\epsilon}^-_{kQ}\sigma_{k}(Q)
 - \Delta_{kQ}^+\phi_k(Q)
 &       + z^-_{kQ} {\bf S}_k(Q)
         - b^+_{kQ} {\bf A}_k(Q)   \\
{[H,\phi_{k}(Q)]} &=&
        - \tilde{\epsilon}^+_{kQ}\psi_{k}(Q)
        - \Delta^+_{kQ}\rho_k(Q)
 &         - b^+_{kQ} {\bf R}_k(Q)
           - z^+_{kQ} {\bf B}_k(Q) \\
{[H,\psi_{k}(Q)]} &=&
        - \tilde{\epsilon}^+_{kQ}\phi_{k}(Q)
        - \Delta^-_{kQ}\sigma_k(Q)
 &         - b^-_{kQ} {\bf S}_k(Q)
           - z^+_{kQ} {\bf A}_k(Q) \\
\end{array}
\label{eq:eom}
\end{equation}
$\Delta_k$,$\tilde{\epsilon}_k$, $b_k$ and $z_k$ were already
defined in the previous section. For the sake of compactness of notation
I introduced
$b^{\pm}_{kQ}\equiv b_{k+Q} \pm b_k$, $z^{\pm}_{kQ}\equiv z_{k+Q} \pm z_k$,
$\Delta^{\pm}_{kQ}\equiv \Delta_{k+Q} \pm \Delta_k$  and
$\tilde{\epsilon}^{\pm}_{kQ}\equiv \tilde{\epsilon}_{k+Q} \pm
\tilde{\epsilon}_k$.\\
The first two terms of all four commutators correspond to
(1) the kinetic energy with exchange self-energy
corrections (Fig. \ref{fig:diagrams}a), and (2) Boguliobov-Valatin
particle-hole mixing (Fig. \ref{fig:diagrams}a'). The remaining two terms
in each of these expressions can be better described with reference to
the definition of the collective coordinates in Eqs. \ref{eq:srab}
and \ref{eq:hi1}. \\
Let us first consider ${\bf R}_k(Q)$ and ${\bf S}_k(Q)$.
The $V(Q)$, $U(Q)$ and $U(k-q)$-terms correspond to the polarization
vertex in the commutators of $\sigma_k$ and $\rho_k$
(Fig. \ref{fig:diagrams}b).
In the commutators of $\phi_k$ and $\psi_k$ the $V(Q)$, $U(Q)$ and
$U(k-q)$-term is a polarization vertex
combined with a particle-hole transformation on one of the legs
(Fig. \ref{fig:diagrams}b'). The $V(k-q)$-terms
correspond to the exchange scattering vertex without (commutators
of $\sigma_k$ and $\rho_k$,Fig. \ref{fig:diagrams}c) and with  particle-hole
transformation
(commutators of $\phi_k$ and $\psi_k$, Fig. \ref{fig:diagrams}c'). \\
Finally ${\bf A}_k(Q)$ and ${\bf B}_k(Q)$ correspond to the direct
particle-particle scattering vertex without (commutators of $\phi_k(Q)$
and $\psi_k(Q)$,
Fig. \ref{fig:diagrams}d) and  with particle-hole
conversion (commutators of $\sigma_k(Q)$ and $\rho_k(Q)$,
Fig. \ref{fig:diagrams}d').\\
If we apply the equations of motion to a general operator of the form
\begin{displaymath}
\hat{O}=\sum_k\left[v_{1,k}(Q)\sigma_k(Q)+v_{2,k}(Q)\rho_k(Q)+
v_{3,k}(Q)\phi_k(Q)+v_{4,k}(Q)\psi_k(Q)\right]
\end{displaymath}
we find that they can be written in matrix form as
\begin{equation}
H^{0}(k,Q) \vec{v}_k(k,Q)
+ \sum_q H^i(k,q,Q)\Gamma(q,Q)\vec{v}_q(Q)
= \nu\vec{v}_k(Q)
\end{equation}
The interaction Hamiltonian $H^i$
contains the matrix elements of Eq. \ref{eq:hi1} on the diagonal, and is
zero elsewhere. We furthermore
use the zero'th order Hamiltonian describing non-interacting
quasi-particles
\begin{equation}
H^{0}(k,Q) \equiv
\left( \begin{array}{cccc}
0                      & \tilde{\epsilon}_{kQ}^{-}      &
   0                      & -\Delta_{kQ}^{-}  \\
\tilde{\epsilon}_{kQ}^{-}      & 0                      &
   -\Delta_{kQ}^{+}      & 0                      \\
0                      & -\Delta_{kQ}^{+} &
   0                      & -\tilde{\epsilon}_{kQ}^{+}     \\
-\Delta_{kQ}^{-}       & 0                      &
   -\tilde{\epsilon}_{kQ}^{+}     & 0                      \\
\end{array}\right)
\end{equation}
and the dimensionless matrix containing coherence factors
\begin{equation}
\Gamma(k,Q) \equiv
\left(\begin{array}{cccc}
0           &z_{kQ}^{-}    &0           &-b_{kQ}^{-}  \\
z_{kQ}^{-}  &0             &-b_{kQ}^{+} &0            \\
0           &-b_{kQ}^{+}   &0           &-z_{kQ}^{+}  \\
-b_{kQ}^{-} &0             &-z_{kQ}^{+} &0            \\
\end{array}\right)
\end{equation}
The collective modes can be found by looking for poles in the correlation
functions, in particular the density-density and the spin-spin correlation
functions $<<T\rho(\vec{r},\tau)\rho(\vec{r'},0)>>_{\nu}$
and $<<T\sigma(\vec{r},\tau)\sigma(\vec{r'},0)>>_{\nu}$, where
$\rho(\vec{r},\tau)$ etc. are the Heisenberg representation of the operators
$\sum_{\vec{Q}} \exp{i\vec{Q}\cdot\vec{r}} \rho(\vec{r})$ which were defined
in the previous section. In the superconducting state also
$<<T\phi(\vec{r},\tau)\phi(\vec{r'},0)>>_{\nu}$ and
$<<T\psi(\vec{r},\tau)\psi(\vec{r'},0)>>_{\nu}$ become relevant.
Together with the six off-diagonal correlation
functions, the four diagonal functions
define a $4\times4$ two-particle Green's function matrix. The matrix
$K^0_{kq}(Q,\nu)=(\nu-H^0(k,Q)-i0^+)^{-1}\delta_{k,q}$
corresponds to the Lehmann representation of this Green's
function in the absence of residual interactions ({i.e.} with $H^i=0$).
As this describes the response of a gas of non-interacting quasi-particles
there are no poles corresponding to collective modes. The generalized
susceptibility $\chi^0(Q,\nu)=\sum_{k,q} \Gamma(k,Q)K^0_{k,q}(Q,\nu)$.
If we now include
$H^i$, we can calculate the Green's functions in the RPA by applying the
Dyson equation
\begin{equation}
K_{kq}(Q,\nu)=K^0_{kq}(Q,\nu)\delta_{k,q}
 +K^0_{kk}(Q,\nu)\sum_{k'}H^i(k,k',Q)\Gamma(k',Q)K_{k'q}(Q,\nu)
\end{equation}
We can use the same partial wave decomposition as introduced in the
previous paragraph where we discussed the gap equation. It is straightforward
to show, that the above Dyson equation has the solution
\begin{equation}
\chi_{\alpha,\beta}(Q,\nu)=
\sum_{\gamma}\chi^0_{\alpha,\gamma}(Q,\nu)
\left(1-H^i \chi^0\right)^{-1}_{\gamma,\beta}
\label{eq:response}
\end{equation}
where I used the partial wave decomposition
\begin{displaymath}
\chi_{\alpha,\beta}(Q,\nu) = \sum_{kq}\psi_{\alpha}(k)
\Gamma(k,Q)K_{kq}(Q,\nu)\psi_{\beta}(q)
\end{displaymath}
with similar expressions for $\chi^0$, and $H^i$.
The collective modes correspond to the zero's of the determinant of
\begin{equation}
\delta_{\alpha,\beta}\delta_{i,j}-\sum_{\mu,l}
H^i_{\alpha,\mu;i,l}  \chi^0_{\mu,\beta;l,j}
\label{eq:determ}
\end{equation}
which can be determined numerically, and in
some limiting cases also analytically. The expression of the response
function Eq. \ref{eq:response} corresponds to calculating the series
of diagrams depicted in Fig. \ref{fig:diagrams}. It is possible to improve
further by taking into account the screening of the vertex in all of these
diagrams, except in the polarization vertices
of Figs. \ref{fig:diagrams}b and
\ref{fig:diagrams}b', as this would lead to double counting of the vertex
corrections. (N.B.: Although in this paper the pairing interaction
is introduced as an independent model
parameter, one should keep in mind, that for an electronic mechanism
of superconductivity such as a spin-fluctuation or plasmon-intermediated
interaction, the pairing arises precisely from these vertex corrections.)
This
procedure was proposed by Anderson, Rickayzen and also by Bardasis and
Schrieffer. Moreover, in the next
section we will see, that in the normal state the
$\sigma$ and $\rho$ channels are completely decoupled for all values
of $Q$. This implies, that the sum over diagrams for the charge fluctuations
does not contain any vertex correction due to the spin fluctuations
and vice-versa. Hence, it is necessary in this case to screen all vertices
in the charge-fluctuation channel with the spin-fluctuations, and
vice versa. As has been shown by Rickayzen, in the superconducting state
the screening properties are basically the same as in the
normal state \cite{rickayzen}. \\
One has to be cautious with this procedure of screening the vertices,
as, by making the RPA {\em before} calculating the sum over diagrams, certain
classes of vertex corrections are omitted. As a result inconsistencies
may arise, as can be seen from
the following example: If we consider the Hubbard U model, the
on-site interaction can
be introduced either using an on-site spin-independent ($V$) or a
singlet-only ($U$) term as defined
in Eq. \ref{eq:hamiltonian}. The expressions for the equation
of motion should be independent of this choice, as the Pauli-exclusion
principle automatically projects out the double occupancy of the same
site with equal spins. Indeed, we can check from Eq. \ref{eq:hi1}
that this requirement is satisfied as long as we do not
introduce screening. If we follow the recipe, that
in the first two lines of Eq. \ref{eq:hi1}
the polarization diagrams $U(Q)$, $V(Q)$, and $U(k-q)$, but not the
exchange diagram $V(k-q)$, should be replaced with the bare interaction,
we arrive at a different result depending on whether we introduce the
on-site interaction through a singlet-only or a spin-independent
interaction. \\
This inconsistency is removed, if we replace
the direct {\em and} exchange terms in $H^i_{\sigma}$ with
the charge screened value. In the same way
screening with spin fluctuations should be introduced 'by hand'
in the direct {\em and} exchange terms in $H^i_{\rho}$.
Finally all three terms in $H^i_{\phi}$ and $H^i_{\psi}$ should be replaced
with the charge {\em and} spin fluctuation- screened vertices.
\\
Let us now calculate $K^0$ by inverting
$\left[\nu-H^{qp}\right]$. The determinant is
\begin{equation}
\begin{array}{l}
\left|\nu - H^{qp}\right|=
\nu^4-2\nu^2(E_{k+Q}^2+E_k^2)+(E_{k+Q}^{2}-E_{k}^{2})^2=\\
(\nu^2-(E_{k+Q}+E_k)^2)(\nu^2-(E_{k+Q}-E_k)^2)
\end{array}
\end{equation}
The zero'th order two-particle Green's function is then
{\scriptsize\begin{equation}
\begin{array}{l}
K^0 =
\left|\nu-H^{qp}\right|^{-1} \times \\
\\
%
\left(\begin{array}{llll}
 \begin{array}{l} \nu(\nu^2-\tilde{\epsilon}^{+^2}-\Delta^{+^2})\\
                                                       \end{array}&
 \begin{array}{l} \tilde{\epsilon}^{-}(\nu^2-\tilde{\epsilon}^{+^2})   \\
   -\tilde{\epsilon}^{+}\Delta^{+}\Delta^{-}                   \end{array}&
 \begin{array}{l} -\nu\tilde{\epsilon}^{-}\Delta^{+}           \\
   +\nu\tilde{\epsilon}^{+}\Delta^{-}                           \end{array}&
 \begin{array}{l} \tilde{\epsilon}^{+}\tilde{\epsilon}^{-}\Delta^{+}   \\
   +(\Delta^{+^2}-\nu^2)\Delta^{-}                       \end{array}\\
&&&\\
 \begin{array}{l} \tilde{\epsilon}^{-}(\nu^2-\tilde{\epsilon}^{+^2})   \\
   -\tilde{\epsilon}^{+}\Delta^{+}\Delta^{-}                   \end{array}&
 \begin{array}{l} \nu(\nu^2-\tilde{\epsilon}^{+^2})            \\
   -\nu\Delta^{-^2}                                    \end{array}&
 \begin{array}{l} -\nu^2\Delta^{+}                     \\
   +(\tilde{\epsilon}^{+}\tilde{\epsilon}^{-}
                      +\Delta^{+}\Delta^{-})\Delta^{-} \end{array}&
 \begin{array}{l} \nu\tilde{\epsilon}^{+}\Delta^{+}            \\
   -\nu\tilde{\epsilon}^{-}\Delta^{-}                          \end{array}\\
&&&\\
 \begin{array}{l} -\nu\tilde{\epsilon}^{-}\Delta^{+}           \\
   +\nu\tilde{\epsilon}^{+}\Delta^{-}                           \end{array}&
 \begin{array}{l} -\nu^2\Delta^{+}                     \\
   +(\tilde{\epsilon}^{+}\tilde{\epsilon}^{-}
                                +\Delta^{+}\Delta^{-})\Delta^{-} \end{array}&
 \begin{array}{l} \nu(\nu^2-\tilde{\epsilon}^{-^2})            \\
   -\nu\Delta^{-^2}                                    \end{array}&
 \begin{array}{l} \tilde{\epsilon}^{+}(\tilde{\epsilon}^{-^2}-\nu^2)   \\
   +\tilde{\epsilon}^{-}\Delta^{+}\Delta^{-}                    \end{array}\\
&&&\\
 \begin{array}{l} \tilde{\epsilon}^{+}\tilde{\epsilon}^{-}\Delta^{+}   \\
   +(\Delta^{+^2}-\nu^2)\Delta^{-}                       \end{array}&
 \begin{array}{l} \nu\tilde{\epsilon}^{+}\Delta^{+}            \\
   -\nu\tilde{\epsilon}^{-}\Delta^{-}                          \end{array}&
 \begin{array}{l} \tilde{\epsilon}^{+}(\tilde{\epsilon}^{-^2}-\nu^2)   \\
   +\tilde{\epsilon}^{-}\Delta^{+}\Delta^{-}                    \end{array}&
 \begin{array}{l} \nu(\nu^2-\tilde{\epsilon}^{-^2}-\Delta^{+^2})
                                                       \end{array}\\
\end{array}\right)
\end{array}
\end{equation}}
%
The $4\times 4$ matrix $K^0\Gamma$ becomes
{\scriptsize\begin{equation}
\begin{array}{l}
K^0\Gamma=|\nu-H^{qp}|^{-1}
\\
\left(\begin{array}{llll}
  \begin{array}{l} -\nu^2(z^-\tilde{\epsilon}^- + b^-\Delta^-) \\
           + (b^-\Delta^+ + z^-\tilde{\epsilon}^+)            \\
   \times(\tilde{\epsilon}^+\tilde{\epsilon}^- +
   \Delta^{+}\Delta^{-})\end{array}&
  \begin{array}{l}z^-\nu^3                              \\
          + \nu z^- (\tilde{\epsilon}^{+^2}+\Delta^{+^2})        \\
       +\nu b^+(\tilde{\epsilon}^+\Delta^--
       \tilde{\epsilon}^-\Delta^+)\end{array}&
  \begin{array}{l} \nu^2(b^+\tilde{\epsilon}^- - z^+\Delta^{-}) \\
           + (z^+\Delta^+ - b^+\tilde{\epsilon}^+)              \\
   \times(\tilde{\epsilon}^{+}\tilde{\epsilon}^{-}+
   \Delta^{+}\Delta^{-})\end{array}&
  \begin{array}{l}    b^-\nu^3                        \\
     +\nu z^+(\tilde{\epsilon}^+\Delta^- - \tilde{\epsilon}^-\Delta^+)  \\
     -\nu b^-(\tilde{\epsilon}^{+^2}+\Delta^{+^2})           \end{array}\\
&&&\\
  \begin{array}{l}z^-\nu^3                              \\
          + \nu z^- (\tilde{\epsilon}^{+^2}+\Delta^{-^2})        \\
       +\nu b^-(\tilde{\epsilon}^+\Delta^+-
       \tilde{\epsilon}^-\Delta^-)\end{array}&
  \begin{array}{l} -\nu^2(z^-\tilde{\epsilon}^- + b^+\Delta^+) \\
           + (b^+\Delta^- + z^-\tilde{\epsilon}^+)            \\
   \times(\tilde{\epsilon}^+\tilde{\epsilon}^-
   + \Delta^{+}\Delta^{-})\end{array}&
  \begin{array}{l}    b^+\nu^3                        \\
     +\nu z^+(\tilde{\epsilon}^+\Delta^+ - \tilde{\epsilon}^-\Delta^-)  \\
     -\nu b^+(\tilde{\epsilon}^{+^2}+\Delta^{-^2})            \end{array}&
  \begin{array}{l} \nu^2(b^-\tilde{\epsilon}^- - z^+\Delta^{+}) \\
           + (z^+\Delta^- - b^-\tilde{\epsilon}^+)              \\
   \times(\tilde{\epsilon}^{+}\tilde{\epsilon}^{-}
   +\Delta^{+}\Delta^{-})\end{array}\\
&&&\\
  \begin{array}{l} \nu^2(-b^-\tilde{\epsilon}^+ + z^-\Delta^{+}) \\
           + (z^-\Delta^- + b^-\tilde{\epsilon}^-)              \\
   \times(\tilde{\epsilon}^{+}\tilde{\epsilon}^{-}
   + \Delta^{+}\Delta^{-})\end{array}&
  \begin{array}{l}        \nu^3b^+                    \\
    -\nu z^-(\tilde{\epsilon}^+\Delta^- - \tilde{\epsilon}^-\Delta^+) \\
          -  \nu b^+(\tilde{\epsilon}^{-^2}+\Delta^{-^2})     \end{array}&
  \begin{array}{l}-\nu^2(z^+\tilde{\epsilon}^+ + b^+\Delta^+)   \\
              +(b^+\Delta^-+z^+\tilde{\epsilon}^-)              \\
   \times(\tilde{\epsilon}^{+}\tilde{\epsilon}^{-}
   + \Delta^{+}\Delta^{-})\end{array}&
  \begin{array}{l}\nu^3z^+                              \\
      -\nu z^+(\tilde{\epsilon}^{-^2}+\Delta^{-^2})             \\
    -\nu b^-(\tilde{\epsilon}^-\Delta^+ -
    \tilde{\epsilon}^+\Delta^-) \end{array}\\
&&&\\
  \begin{array}{l}    b^-\nu^3                        \\
     -\nu z^-(\tilde{\epsilon}^+\Delta^+ - \tilde{\epsilon}^-\Delta^-)  \\
     -\nu b^-(\tilde{\epsilon}^{-^2}+\Delta^{+^2})            \end{array}&
  \begin{array}{l} -\nu^2(b^+\tilde{\epsilon}^+ - z^-\Delta^-)\\
           - (z^-\Delta^+-b^+\tilde{\epsilon}^-)              \\
   \times(\tilde{\epsilon}^{+}\tilde{\epsilon}^{-}
   + \Delta^{+}\Delta^{-})\end{array}&
  \begin{array}{l}z^+\nu^3                              \\
          - \nu z^+ (\tilde{\epsilon}^{-^2}+\Delta^{+^2})        \\
       +\nu b^+(\tilde{\epsilon}^+\Delta^+
       -\tilde{\epsilon}^-\Delta^-)\end{array}&
  \begin{array}{l}-\nu^2(z^+\tilde{\epsilon}^+ + b^-\Delta^-)   \\
       +(z^+\tilde{\epsilon}^- + b^-\Delta^+)                   \\
   \times(\tilde{\epsilon}^{+}\tilde{\epsilon}^{-}
   + \Delta^{+}\Delta^{-})\end{array}\\
\end{array}\right)
\end{array}
\end{equation}}
{}From inspection of the matrix elements it turns out, that
they all contain the factor $(\nu^2-(E_{k+Q}-E_k)^2)$ in the
numerator. As the same term appears in the denomenator, these factors
cancel. $K^0\Gamma$ turns out to be symmetric, and the exact result is
{\scriptsize\begin{equation}
\begin{array}{l}
\Gamma K^0 = K^0\Gamma =
\frac{E_{k+Q}+E_k}{2E_kE_{k+Q}((E_{k+Q}+E_k)^2-\nu^2)} \\
\\
\left(\begin{array}{cccc}
\begin{array}{c}-E_{k+Q}E_k+\Delta_k\Delta_{k+Q}\\
+\tilde{\epsilon}_k\tilde{\epsilon}_{k+Q}\end{array}&
\nu\frac{E_{k+Q}\tilde{\epsilon}_k-E_k\tilde{\epsilon}_{k+Q}}{E_{k+Q}+E_k} &
\tilde{\epsilon}_{k+Q}\Delta_k-\tilde{\epsilon}_k\Delta_{k+Q}   &
-\nu\frac{E_{k+Q}\Delta_k-E_k\Delta_{k+Q}}{E_{k+Q}+E_k}   \\
&&&\\
\nu\frac{E_{k+Q}\tilde{\epsilon}_k-E_k\tilde{\epsilon}_{k+Q}}{E_{k+Q}+E_k}   &
\begin{array}{c}-E_{k+Q}E_k-\Delta_k\Delta_{k+Q}\\
+\tilde{\epsilon}_k\tilde{\epsilon}_{k+Q}\end{array}&
\nu\frac{E_{k+Q}\Delta_k+E_k\Delta_{k+Q}}{E_{k+Q}+E_k}       &
-\tilde{\epsilon}_{k+Q}\Delta_k-\tilde{\epsilon}_k\Delta_{k+Q}     \\
&&&\\
\tilde{\epsilon}_{k+Q}\Delta_k-\tilde{\epsilon}_k\Delta_{k+Q}    &
\nu\frac{E_{k+Q}\Delta_k+E_k\Delta_{k+Q}}{E_{k+Q}+E_k}       &
\begin{array}{c}-E_{k+Q}E_k-\Delta_{k+Q}\Delta_k\\
-\tilde{\epsilon}_{k+Q}\tilde{\epsilon}_k\end{array}&
\nu\frac{E_{k+Q}\tilde{\epsilon}_k+E_k\tilde{\epsilon}_{k+Q}}{E_{k+Q}+E_k} \\
&&&\\
-\nu\frac{E_{k+Q}\Delta_k-E_k\Delta_{k+Q}}{E_{k+Q}+E_k}    &
-\tilde{\epsilon}_{k+Q}\Delta_k-\tilde{\epsilon}_k\Delta_{k+Q}               &
\nu\frac{E_{k+Q}\tilde{\epsilon}_k+E_k\tilde{\epsilon}_{k+Q}}{E_{k+Q}+E_k}   &
\begin{array}{c}-E_{k+Q}E_k+\Delta_k\Delta_{k+Q}\\
-\tilde{\epsilon}_k\tilde{\epsilon}_{k+Q}\end{array}\\
\end{array}\right)
\end{array}
\label{eq:G-final}
\end{equation}}
\section{Examples}
In this section I will apply the formalism outlined above to
a number of examples with an increasing degree of complexity {\em
vis a vis} the type of electron-electron interaction that is assumed.
The energy dispersion is assumed to be of the form
\begin{equation}
\epsilon_k=-2t\left(\cos(k_xa)+\cos(k_ya)\right)
   -2t'\cos(k_xa)\cdot\cos(k_ya) - \mu
\label{eq:dispersion}
\end{equation}
where $a$ $b$ and $c$ are the lattice parameters.
The $t$ and $t'$-terms are due to nearest-neighbour and next-nearest
neighbour hopping in a square lattice. If $t'=0$ at half filling
of the band, such a dispersion relation has the remarkable property
that the Fermi surface forms a perfect square, with a diverging
effective mass over the entire Fermi surface. In
practice this situation will never occur, as there will always be
some finite coupling between next nearest neighbours. This causes
a bulging of the Fermi surface, which eventually transforms into a rotated
Fermi surface if $|t'| \gg |t|$.
\\
In all examples I will restrict the discussion to systems where
electrons have an on-site attraction or repulsion, a nearest-neighbour
interaction, or both, as well as the long-range $e^2/r$
repulsive interaction. Moreover the discussion is limited to the
situation where a single band crosses the Fermi surface, and
tight-banding language will be used for the description of this
band. In particular I will consider a tightbinding band on a
three-dimensional square lattice, with a strong anisotropy
leading to quasi two-dimensional behaviour. A convenient set of
functions to be used for the partial-wave decomposition of $H^i$
is then the set of harmonic functions:
\begin{equation}
\begin{array}{llll}
s\mbox{:}          &\psi_0(k)&=& 1                        \\
s^*\mbox{:}        &\psi_1(k)&=& \cos(k_x a)+\cos(k_y b)  \\
d_{x^2-y^2}\mbox{:}&\psi_2(k)&=& \cos(k_x a)-\cos(k_y b)  \\
p_{x}\mbox{:}      &\psi_3(k)&=& \sqrt{2} \sin(k_x a)     \\
p_{y}\mbox{:}      &\psi_4(k)&=& \sqrt{2} \sin(k_y b)     \\
d_{xy}\mbox{:}     &\psi_5(k)&=& 2 \sin(k_x a) \sin(k_y b)\\
{\mbox  etc.}  &&&\\
\end{array}
\end{equation}
The $k$-space representation of the on-site Hubbard $U$ interaction
$\sum_i U_0 n_{i\uparrow}n_{i\downarrow}$
is the $k$-independent function $U(k-q)=U_0\psi_0(k)\psi_0(q)$.
If we consider the nearest-neigbour interaction
$ \frac{1}{2}V_{1}\sum_{<i,j>} n_i n_j $
we find, by means of a direct Fourier transformation of the operators
$n_i=c^{\dagger}_{i\uparrow} c_{i\uparrow}
+ c^{\dagger}_{i\downarrow} c_{i\downarrow}$,
that this can be cast in the form
$\frac{1}{2}\sum_Q V(Q) \rho(Q)\rho(-Q)$
with $V(Q)= 2V_{1}\psi_1(Q)$, so that we
obtain the partial-wave decomposition
\begin{equation}
V(k-q)=V_1\left(\psi_1(k)\psi_1(q)+\psi_2(k)\psi_2(q)
-\psi_3(k)\psi_3(q)-\psi_4(k)\psi_4(q)\right)
\end{equation}
A singlet-only nearest-neigbour interaction
$\frac{1}{4}U_1\sum_{<i,j>}
(c^{\dagger}_{i\uparrow}c^{\dagger}_{j\downarrow}
-c^{\dagger}_{i\downarrow}c^{\dagger}_{j\uparrow})
(c_{j\downarrow}c_{i\uparrow}-c_{j\uparrow}c_{i\downarrow})$
can be cast in the form
$\frac{1}{8}\sum_{Q}U(Q)\left[ \rho(Q) \rho(-Q)
 - \vec{\sigma}(Q) \cdot \vec{\sigma}(-Q) \right]$ with
$U(Q)=2U_{1}\psi_1(Q)$, hence it has the same partial wave
expansion as $2V_{1}\psi_1(Q)$.
However, from Eq. \ref{eq:hi1} we see, that the
singlet-only interaction has other prefactors, and is summed over
$U(k-q)$ and $U(k+q)$ in the pairing channel. \\
Finally we have to take into account the long range Coulomb interaction.
Here we will use the lattice Fourier transform of $e^2/r$.
The screening of the Coulomb interaction for part of the
vertices has been discussed above, and is essential, as a
bare $Q^{-2}$ interaction is known to create a singularity at
the Fermi level within the random phase approximation.
I will use the convention in the remainder of this
paper, that $V(Q)$ is the bare Coulomb repulsion at large
distances, whereas for shorter distances $U_0$ and $V_1$ are the
projections of $V(Q)$ on the on-site interaction and
the spin-independent nearest neighbour interaction respectively.
Taking all these terms together we obtain for a model with
a 'singlet-only' nearest neighbour interaction
\begin{equation}
\begin{array}{llll}
H^i_{\sigma}(0,0) &=&         - U_0^{\rho}   -
U_1^{\rho}\psi_1(Q)   &        \\
H^i_{\sigma}(\alpha,\alpha) &=& - U_1^{\rho}/2
& (\alpha=1,2)\\
H^i_{\rho}(0,0)   &=& 2V^{\sigma}(Q)  - U_0^{\sigma}
+ U_1^{\sigma}\psi_1(Q) &\\
H^i_{\rho}(\alpha,\alpha)   &=&   U_1^{\sigma}/2
& (\alpha=1,2)\\
H^i_{\phi}(0,0)   &=&           U_0^{\rho\sigma}
&        \\
H^i_{\phi}(\alpha,\alpha)   &=&   U_1^{\rho\sigma}
& (\alpha=1,2)\\
\end{array}
\label{eq:hi2}
\end{equation}
For all symmetries we have
$H^i_{\psi}(\alpha,\beta)=H^i_{\phi}(\alpha,\beta)$.
The upper-indices $\rho$ and $\sigma$ indicate whether screening with
charge- or spin-fluctuations is implied.  The minus-sign in front
of the $U_0$ term in $H^i_{\rho}(0,0)$ is not a misprint. As $2V(Q)$,
'contains' the on-site Hubbard term, the sum of these two contributions
is $+U_0$. In principle one should
also include higher harmonics, as the expansion of $V(Q)$ does not end at
$\psi_4$. However, as the expansion only appears as a {\em screened}
interaction in expression \ref{eq:hi2}, it is reasonable to work with a
a model where such interaction-terms are neglected.
If the nearest-neigbour interaction is spin-independent
we must also include $p_x$ and $p_y$ symmetries of pairing, and we obtain
\begin{equation}
\begin{array}{llll}
H^i_{\sigma}(0,0) &=&         - U_0^{\rho}             &          \\
H^i_{\sigma}(\alpha,\alpha) &=&         - V_1^{\rho}   & (\alpha=1..4) \\
H^i_{\rho}(0,0)   &=& 2V^{\sigma}(Q)  - U_0^{\sigma}   &          \\
H^i_{\rho}(\alpha,\alpha)   &=&         - V_1^{\sigma} & (\alpha=1..4) \\
H^i_{\phi}(0,0)   &=&           U_0^{\rho\sigma}       &          \\
H^i_{\phi}(\alpha,\alpha)   &=&       V_1^{\rho\sigma} & (\alpha=1..4) \\
\end{array}
\label{eq:hi3}
\end{equation}
In the previous section we have seen, that in addition to the partial-wave
expansion of $H^i$, we also have to make a similar expansion of
$\Gamma K^0$. The expression for this product is
given in Eq. \ref{eq:G-final}. The partial-wave expansion of this expression
is in general complicated, and has to be done with the help of
a computer. Some limiting cases exist however where the integrals
can be solved, especially when an expansion for small $Q$ can be
made. Some of these limiting cases will be treated in the subsequent
sections. In addition numerical calculations will be given at
general values of the collective mode momentum $Q$.
\subsection{Normal state limit}
In the non-superconducting limit Eq. \ref{eq:G-final} has only non-vanishing
matrix elements on the diagonal. Furthermore only the charge and spin-channels
are relevant in the absence of off-diagonal order.
Let us make the further assumption that the electrons interact with each other
via an on-site Hubbard $U$ repulsion, which is therefore independent of $k$.
After the summation over $k$ we obtain for the top-left corner
of Eqs. \ref{eq:G-final}
\begin{equation}
\begin{array}{l}
H^i(Q)\chi^0 =
\sum_k\frac{(\epsilon_{k+Q}-\epsilon_{k})(f_{k}-f_{k+Q})}
           {\nu^2-(\epsilon_{k+Q}-\epsilon_{k})^2}
\left(\begin{array}{cc}
 -U_0^{\rho}   &
 0        \\
 0         &
 2V^{\sigma}(Q)-U_0^{\sigma}   \\
\end{array}\right)
\end{array}
\end{equation}
where the $f_k$ are Fermi occupation factors. Let us assume for this part of
the discussion, that the momentum of the electrons in the plane is unbounded.
In that case the Fourier transform of $e^2/r$ is discrete in the
direction perpendicular to the planes, and continuous along the planes,
so that\cite{fetter}
$V(Q)=2\pi e^2d^2 \sinh{(Q_{\parallel}d)}
\left[Q_{\parallel}d\right]^{-1}
\left[\cosh{(Q_{\parallel}d)}-\cos{(Q_{\perp}d)}\right]^{-1}$
where $d$ is the interlayer distance.
Let us define $\tilde{Q}=\sqrt{4\pi e^2 /V(Q)}$, which has
the property $\lim_{Q\rightarrow 0} \tilde{Q} = Q$. If we assume that we have
a cylindrical Fermi surface, with an isotropic Fermi velocity $v_F$, and a
Fermi wavevector $k_F$, we obtain with
$1-\epsilon=2V(Q)\chi^0(2,2)$, and $\nu_p^2\equiv 2e^2d^{-1}\hbar k_Fv_F$
\begin{equation}
\begin{array}{lll}
\epsilon&=&1-
   2V(Q) \frac{2\pi k_F}{d} \frac{(2\pi)^{-3}}{\hbar v_F}
   \int_0^{2\pi}d\phi
   \frac{\hbar^2Q_{\parallel}^2v_F^2\cos^2\phi}
   {\nu^2-\hbar^2Q_{\parallel}^2v_F^2\cos^2\phi} \\
                 &=&1-\frac{2\nu_p^2}{\hbar^2\tilde{Q}^2v_F^2} \left\{
   \left(1-\frac{\hbar^2Q_{\parallel}^2v_F^2}{\nu^2}\right)^{-1/2}
   - 1 \right\}                 \\
\end{array}
\end{equation}
The plasma dispersion relation becomes
\begin{equation}
\nu=\nu_p\frac{Q_{\parallel}}{\tilde{Q}}
     \frac{1+\hbar^2\tilde{Q}^2v_F^2/(2\nu_p^2)}
     {\left\{1+\hbar^2\tilde{Q}^2v_F^2/(4\nu_p^2)\right\}^{1/2}}
 =\nu_p\frac{Q_{\parallel}}{\tilde{Q}}
 \left\{
     1+\frac{3}{8}\frac{\hbar^2\tilde{Q}^2v_F^2}{\nu_p^2} + ...
 \right\}
\end{equation}
The spin-susceptibility per unit cell
($\Omega_u$ is the area in the 2D plane) is
\begin{equation}
\chi(1,1)=\frac{(1-\hbar^2Q_{\parallel}^2v_F^2/\nu^2)^{-1/2}-1}
         {W+U_0[(1-\hbar^2Q_{\parallel}^2v_F^2/\nu^2)^{-1/2}-1]}
\end{equation}
where $W\equiv\frac{m\Omega_u}{\hbar^2\pi}$ is the effective bandwidth.
We see, that in the high frequency limit ($\nu \gg \hbar Q_{\parallel} v_F$)
$\chi^{HF}(1,1)=E_FQ_{\parallel}^2
\Omega_u[\nu^2+U_0E_FQ_{\parallel}^2 \Omega_u]^{-1}$,
and in the low frequency limit $\chi^{LF}(1,1)=(U_0-W)^{-1}$,
Hence the AC susceptibility is suppressed, whereas the static
susceptibility is enhanced. A magnetic instability occurs for $U_0\approx W$.
The above expressions are derived assuming
a free electron dispersion. If the Fermi surface
has nesting vectors\cite{scalapino},
instabilities for specific values of $Q$ are
often found.
\subsection{s-wave superconductivity}
For s-symmetry, and neglecting the radial $k$-dependence of the
pairing potential, the partial wave decomposition of $\chi^0$ and $H^i$
is trivially achieved by summing over all $k$. As a model for the
pairing-interaction we adopt $U(k-q)=-g$. As is ususally done in the
gap-equation, one can limit the energy-range of the interactions
in these expressions by putting $\Delta=0$ for energies larger then a
scaling value (the Debije frequency for phonon-mediated pairing). For the
long range Coulomb interaction we take again $V(Q)=4\pi e^2 Q^{-2}$.
Due to the fact, that $\epsilon_{k+Q}=\epsilon_{-k-Q}$ we obtain
after summation that $\chi^0(1,2)=\chi^0(1,3)=\chi^0(1,4)=0$.
Hence the spin-fluctuations are fully decoupled from the other three
and can be considered separately.
The remaining diagonal and off-diagonal susceptibilitues are finite,
and the following expressions are required
\begin{equation}
\begin{array}{lll}

\chi^0(1,1)&=&\sum_k
       \frac{\left(E_{k}+E_{k+Q}\right)
    \left(E_{k}E_{k+Q}-\Delta_k\Delta_{k+Q}-\epsilon_{k}\epsilon_{k+Q}\right)}
            {2E_{k}E_{k+Q}\left(\nu^2-\left[E_{k}+E_{k+Q}\right]^2\right)} \\

\chi^0(2,2)&=&\sum_k
       \frac{\left(E_{k}+E_{k+Q}\right)
    \left(E_{k}E_{k+Q}+\Delta_k\Delta_{k+Q}-\epsilon_{k}\epsilon_{k+Q}\right)}
            {2E_{k}E_{k+Q}\left(\nu^2-\left[E_{k}+E_{k+Q}\right]^2\right)} \\

\chi^0(3,3)&=&\sum_k
       \frac{\left(E_{k}+E_{k+Q}\right)
    \left(E_{k}E_{k+Q}+\Delta_k\Delta_{k+Q}+\epsilon_{k}\epsilon_{k+Q}\right)}
            {2E_{k}E_{k+Q}\left(\nu^2-\left[E_{k}+E_{k+Q}\right]^2\right)} \\

\chi^0(4,4)&=&\sum_k
       \frac{\left(E_{k}+E_{k+Q}\right)
    \left(E_{k}E_{k+Q}-\Delta_k\Delta_{k+Q}+\epsilon_{k}
    \epsilon_{k+Q}\right)}
            {2E_{k}E_{k+Q}\left(\nu^2-\left[E_{k}+E_{k+Q}\right]^2\right)} \\

S&=&\sum_k
       \frac{\left(E_{k}+E_{k+Q}\right)}
            {2E_{k}E_{k+Q}\left(\nu^2-\left[E_{k}+E_{k+Q}\right]^2\right)} \\

T&=&\sum_k
       \frac{\left(E_{k}+E_{k+Q}\right)
             \left(\epsilon_{k}+\epsilon_{k+Q}\right)}
            {2E_{k}E_{k+Q}\left(\nu^2-\left[E_{k}+E_{k+Q}\right]^2\right)} \\

M&=&\sum_k
       \frac{\left(E_{k}+E_{k+Q}\right)
             \left(\epsilon_{k}-\epsilon_{k+Q}\right)^2}
            {2E_{k}E_{k+Q}\left(\nu^2-\left[E_{k}+E_{k+Q}\right]^2\right)} \\

N&=&\sum_k
       \frac{\left(E_{k}-E_{k+Q}\right)
             \left(\epsilon_{k+Q}-\epsilon_{k}\right)}
            {2E_{k}E_{k+Q}\left(\nu^2-\left[E_{k}+E_{k+Q}\right]^2\right)} \\

\end{array}
\end{equation}

Let us first consider the spin susceptibility. As now
$\chi(1,1)=\frac{\chi^0(1,1)}{1-g\chi^0(1,1)}$ with $g >0 $ for a
BCS interaction,
the spin-fluctuations are pushed to a slightly higher energy.

The generalized susceptibility for density, phase and amplitude
can be expressed using the above definitions as
\begin{displaymath}
\left(\begin{array}{ccc}
\chi^0(1,1)+2\Delta^2S & -\nu\Delta S & \Delta T\\
&&\\
-\nu\Delta S & \chi^0(3,3) & -\nu(T+N)/2\\
&&\\
\Delta T &-\nu (T+N)/2 & \chi^0(4,4)\\
\end{array}\right)
\end{displaymath}

Using the gap equation, ($1=\sum_k\frac{g}{2E_k}$) it is easy to prove, that
$\chi^0(3,3)=-1/g+\nu^2S/2-M/2$ and $\chi^0(4,4)=\chi^0(3,3)-2\Delta^2S$.
Using these properties, we see that the matrix $1-H^i\chi^0$ becomes
\begin{displaymath}
\left(\begin{array}{ccc}
1-2V_Q\chi^0(1,1)-4V\Delta^2S & 2V_Q\nu\Delta S & -2V_Q\Delta T\\
&&\\
-g\nu\Delta S & g(\nu^2S-M)/2 & -g\nu(T+N)/2\\
&&\\
g \Delta T & -g\nu (T+N)/2 & g([\nu^2-4\Delta^2]S-M)/2 \\
\end{array}\right)
\end{displaymath}
and the collective modes can be calculated from the determinant
\begin{equation}
\begin{array}{lll}
0&=&(1-2V_Q\chi^0(1,1))
  \left(
    \left(\nu^2S-M\right)
    \left([\nu^2-4\Delta^2]S-M\right)
    -\nu^2(T-N)^2
  \right) \\
&&+ 4V\Delta^2
  \left(
       (N^2+2TN)S\nu^2+M(S^2[\nu^2-4\Delta^2]-T^2
  \right)
\end{array}
\end{equation}
To further analyze this expression we need to make a series expansion
for small $Q$. In what follows we will neglect
$N$ ($\propto Q^2$, but with a vanishing prefactor if the gap has
electron hole symmetry). Furthermore we notice, that we can
write $2V_Q\chi^0(1,1)\approx\nu_p(Q)^2(\nu^2-W^2-4\Delta^2)^{-1}$,
$M\approx S<(\vec{v_F}\cdot\vec{Q})^2>$, $V_Q M\approx
(1+W^2/4\Delta^2)V_Q\chi^0(1,1)$,
and $T\approx-2\mu_FS$, where $\mu_F=E_F-W/2$, and $W$ is the
effective bandwidth. Retaining
only leading orders in $<(\vec{v_F}\cdot\vec{Q})^2>$ we see, that the
collective modes can now be solved from
\begin{equation}
2(\nu^2-W^2-4\Delta^2-\nu_p(Q)^2)(\nu^2+2\Delta^2)
<(\vec{v_F}\cdot\vec{Q})^2>
=(\nu^2-4\Delta^2-4\mu_F^2)(\nu^2-W^2-4\Delta^2)(\nu^2-\nu_p(Q)^2)
\end{equation}
Hence we see, that the negative $U$ Hubbard model permits
in principle four collective modes: a
spin-density oscillation discussed above, a plasma-mode,
and two additional modes, which are however situated in the
two-quasiparticle continuum, and therefore are strongly damped.
Interestingly the plasma-like mode can exist at frequencies above
{\em} and below the gap, depending on the initial value of
$\nu_p(Q)$ in the normal state. As has been discussed by
Fertig and das Sarma \cite{fertig} a layer dispersion
relation as discussed above, permits the existance of low
lying plasmons {\em below} the gap. Another mechanism for
reducing the plasma frequency in the superconducting state
is strong damping of the motion perpendicular to the planes,
as we recently discussed \cite{damping}.
\subsection{s-wave superconductivity in a layered electron gas.}
If the superconductor is strongly anisotropic, the plasma energy for
$Q\rightarrow 0$ depends on the direction of propagation. An extreme
example of this arises, when the mass in one of the three
directions is infinite,
resulting in a system which behaves two-dimensional from
the point of view of the signle particle band-structure,  whereas the
Coulomb forces are three dimensional. A simple
model exhibiting such behaviour is an infinite stack of a
two-dimensional layers.
The electrodynamics of this system was already
discussed by several authors\cite{fetter,bose,morawitz} using
hydrodynamic calculations, as well as with the random phase approximation.
The resulting plasmon spectrum of such a metal is, in the limit of a large
wavelength, $\nu(\vec{Q})=\nu_p Q_{\parallel}|\vec{Q}|^{-1}$, which
for $Q_{\perp}=0$ saturates at the value
$\nu_p$, while for finite values of  $Q_{\perp}$ it has an
accoustic-like dependence on $Q_{\parallel}$. \\
This implies that here we have a system which on the one hand has
a density of charge carriers characteristic of a metal, and,
provided that there is a pairing mechanism, therefore
has the potential of becoming a BCS-like superconductor. On the other hand
the dynamical response of the electrons in one of the
directions is more characteristic
of a semiconductor or an insulator. This combination
provides us with an example
where the Anderson-Higgs mechanism does {\em not} shift the Goldstone
mode to a high energy, in spite of the fact that the
particles interact through a long
range Coulomb force. Here we will use the dispersion introduced in
Eq. \ref{eq:dispersion} with $t'=0$.
In this example $W=4t$ is the bandwidth. For the long
range Coulomb forces we take the lattice Fourier transform of
$e^2/r$,
which has $\Omega_u^{-1} 4\pi e Q^{-2}$ as its long
wavelength limiting behaviour,
and posesses the same periodicity in $k$-space as the tight-binding band.
$\mbox{Ry}^*=e^2d^{-1}$ is the effective Rydberg which,
together with the Fermi energy, sets the scale of the
plasma frequency in the planar direction
($\nu_p = 2\sqrt{E_F \mbox{Ry}^*}$ for
$E_F<<W$). For the pairing interaction we adopt the same model as
in the previous section, with the following set of
parameters: $E_F/W=0.35$, $g/W=0.6729$ (resulting in $\Delta/W = 0.25$), and
$\mbox{Ry}^*/W=4.0$. In Fig. \ref{fig:layerdispersion}a
the result is displayed with
$Q_{\perp}$ as a parameter in the range from $0$ to $d^{-1}\pi$, and agrees
well with the calculations of Fertig and das Sarma, and those
by Cote and Griffin\cite{fertig,cote}. Due to the model assumption of
an energy independent attraction, the electron-hole continuum becomes
a broad band already for zero momentum in the two-particle channel. In
Fig. \ref{fig:layerdispersion}b  the same set of
calculations is displayed for the normal
state. The plasma frequencies become somewhat smaller in the
superconducting state, which is due to the fact that
the gap in this example is relatively large. It reflects a well-known
property of the negative $U$ Hubbard model, that the mass of a pair is
enhanced due to the fact that two particles have to
hop simultaneously\cite{micnas2}. Also a
strong qualitative difference arises, which is not directly evident
from these curves. This is the change in character of the modes.
In Fig. \ref{fig:layerdispersion}c
the distribution of weight of the mode over the density, gap-phase, and
gap-amplitude branches is displayed.
First of all we notice that the contribution
of the latter is neglegible. The second
interesting feature is, that the nature
of the collective mode changes gradually
from a pure phase-fluctuation at low energy to
a 50/50 percent phase-density mixture at the edge of the
particle-hole continuum. Inside the
electron-hole continuum the collective modes are damped.
(Although they may still
persist as a resonance, they can not be
indentified from the zero's of a determinant)
However, for energies larger than the
particle hole continuum of our band we see
that the density-fluctuation character dominates.
\\
The plasmon-dispersion, which is coupled to the $s$-phase
fluctuating channnels
for small $q$, is accoustic for finite $Q_{\perp}/Q_{\parallel}$
so the Landau criterion \cite{holyscripture} remains satisfied
in spite of having a gapless plasmon spectrum.
\subsection{From s- to d-wave superconductivity in a layered electron gas.}
Let us consider a spin-dependent nearest-neighbour
pairing interaction ($U_1$). As a result the only
channels open for pairing are s and d-wave, and these are also the
only channels in which a bound state can be pulled below the electron-hole
continuum. Let us furthermore assume the same bandstructure as in the
previous section. The phase-diagram, displayed in Fig. \ref{fig:phase1}
was calculated by
searching numerically for the minumum of the Helmholz free energy
(Eq.\ref{eq:rickomega} + $\mu N_e$, using $\mu$ as a Lagrange
parameter to keep the number of electrons fixed) as a function
of $\Delta_s$, $\Delta_{s^*}$ and $\Delta_{d(x^2-y^2)}$. The boundaries,
which are indicated in this diagram are calculated for
$T=0.01W$, where $W=4t$ is the bandwidth. For $T=0$ no sharp phase
boundaries exist. Somewhat
surprisingly, for $|U_1|$ larger than a critical value (which depends
on $n_e$), the ground state is of mixed $\alpha s + \beta d$ symmetry,
which is automatically a state of broken spatial symmetry. However
the strong interaction which is required probably does not exist
in any realistic model of superconductivity. It is worthwhile to
mention in this context, that the region of $sd$-mixing almost
coincides with the region of $p$-wave symmetry, if we use a spin-independent
interaction ($V_1$) instead.
\\
Let us now consider the collective modes for the examples along the line AB
indicated in the phase diagram.
The result is displayed in Fig. \ref{fig:softphason}.
We see, that a soft-mode developes if we approach the phase boundary between
$s$ and $d$-wave superconductivity. The transition takes place exactly when
the mode has developed into a Bogoliubov sound-mode. If we keep imposing the
$s$-symmetry for the ground-state, while actually being in the d-wave part
of the phase-diagram, we always find a soft-mode of phase-fluctuating
character, indicating that the solution is instable. If we allow
the groundstate wave-function to become $d$-wave paired, the gap disappears,
and a sound-wave phase-fluctuation mode occurs directly below the
particle-hole continuum. Both the Bogoliubov
mode and the lower bound of the particle-hole continuum are sound-like, so
that according to Landau's argument \cite{holyscripture} a
supercurrent-flow is still possible in spite of the fact that there is no gap.
\subsection{Phase versus spin-fluctuating modes in a layered electron gas.}
Let us now consider the singlet-only nearest-neighbour
pairing interaction $U_1$.
In the discussion of the resonating valence bond state
\cite{bza1,bza2,kotliar}
the t-J model has been used, where $J=-U_1$, and a reduction of the
double occupancy of the same site is included, either by replacing the
bare hopping parameter $t$ with an effective one, or by using more elaborate
schemes. It is not the aim of the present discussion to
address the $tJ$ model.
Instead we consider a Fermi-liquid, with an
on-site repulsion ($U_0$) which is not
too strong, and an attractive interaction between electrons on
a neighbouring site ($U_1$). As the actual bandstructure
in these systems is experimentally known to be better described
by the three band model of Zaanen, Sawatzky and Allen\cite{zsa},
(which is again a simplified version of the real valence band structure
involving 6 oxygen 2p bands and 5 cupper 3d-bands for the occupied
states, as well as unoccupied 3s and 3p states)
a transformation to a single band hamiltonian will in principle
generate both an effective Hubbard $U_0$ and an
intersite $U_1$\cite{zr,es,jef}. Examples of such
transformations can be found
in the work by Emery \cite{emery}, and by Jansen \cite{jansen}.
However, also other, more complicated types of
interactions are generated when
making such transformations, notably the correlated hopping term (with
six operators) which, as has been shown by Hirsch, promotes
superconductivity of hole-carriers\cite{hirsch}. The interaction
considered by Jansen as well as the correlated
hopping term treated by Hirsch,
effectively provide an {\em on-site} attraction,
which, when considered on its own, promotes pairing in the (non-extended)
s-wave channel.
Also $U_1$ term contains contributions from the virtual exchange
of spin-fluctuations\cite{monthoux92,monthoux93}. As has been discussed by
Scalapino\cite{scalapino}, such processes give
rise to an attraction on nearest neighbour sites, and increase the
on-site repulsion between electrons. As the exchange spin-fluctuations are
really vertex corrections due to the $H^i_{\sigma}$-channel, one could
schematically regard $U_1$ in Eq. \ref{eq:hi2} as the vertex correction
of $U_0$. As such corrections are necessarily retarded,
and therefore rather ill-represented by the non-retared
interaction assumed here, the present analysis can at best provide
a qualitative picture.
\\
BZA\cite{bza2} considered pairing of the $s^*$-type
near half filling, Emery considered $d_{x^2-y^2}$-pairing,
and Kotliar studied both $s^*$ and $d$-type pairing.
As we will see, the $s^*$-type pairing is not a stable
solution near half filling, and is dominated by pairing of the $d$-type.
As the latter again tends to be unstable with respect to the
anti-ferromagnetic Mott-Hubbard insulating state at half filling,
superconductivity can only exist sufficiently
far away from this region. As the optimal $T_c$ would
have been reached at half filling for a symmetrical band, this
would lead to the conclusion that superconductivity is only
a marginal effect in such a system. However, the high $T_c$ cuprates
do not have an electron-hole symmetrical band, and the Fermi
surface is known to be strongly
distorted from the perfect square that arises from considering
only nearest neighbour hopping. This actually comes to rescue:
As a function of band-filling it pulls apart the regions, where
anti-ferromagnetism and high $T_c$ have their highest stablility.\\
The three coupled gap equations are (with $x\equiv k_xa$ and $y\equiv k_ya$)
\begin{equation}
\left(1+\sum_k\frac{\tanh{E_k/(2k_BT)}}{2 E_k}
\left[\begin{array}{ccc}
U_0               &U_0[\cos x+\cos y]  &U_0[\cos x-\cos y]  \\
U_1[\cos x+\cos y]&U_1[\cos x+\cos y]^2&U_1[\cos^2x-\cos^2y]\\
U_1[\cos x-\cos y]&U_1[\cos^2x-\cos^2y]&U_1[\cos x-\cos y]^2\\
\end{array}\right]\right)
\left(\begin{array}{c}
\Delta_s\\ \Delta_{s^*} \\  \Delta_{d}
\end{array}\right) = 0
\label{eq:gapd}
\end{equation}
together with a fourth expression, which determines the chemical potential
by constraining the electron-occupation number
$\sum_k (1-\epsilon_k/E_k)=N_e$.
For a sufficiently small value of $U_1$ or for $T$ near $T_c$, where $\Delta$
becomes small, the denomenator has the four-fold symmetry of the crystal,
and the
cross-terms linking $s^*$ to $d$ are zero for symmetry reasons. Hence only
$s$ and $s^*$ are coupled provided that $U_0 \neq 0$.
If $\Delta$ becomes large compared to the bandwidth, {\em a priori}
there is no reason why mixing between $s$ and $d$ is forbidden, and indeed
we will see, that such a mixing takes place for a large value of $U_1$.
\\
I still need to specify the electron dispersion relation before
we can solve the gap equations. For the dispersion relation we now
use Eq. \ref{eq:dispersion} with $t'=-0.7t$. The shape of the
Fermi surface obtained with this choice of parameters
is very close to what has been
calculated with the local density approximation for {\em e.g.}
La$_2$CuO$_4$ and YBa$_2$Cu$_3$O$_7$ \cite{pickett,shen}. Due to
the finite value of $t'$ a significant
change occurs in the density of states (DOS) at the Fermi energy
as a function of the
number of electrons per unit cell. The DOS is now a-symmetric, and the
maximum is shifted to the 'hole-doped' side of the point where
the band is half filled. Of course the
direction in which this occurs is dictated by the sign of $t'$. With
$t'<0$ we mimic the situation encountered in
the $CuO_2$-planes of the high $T_c$ cuprates. \\
\\
The phase diagram with $t'/t=-0.7$ and $U_1/(4t)=-0.5$, and $U_0/(4t)=0$
is displayed in Fig. \ref{fig:phase2}.
Due to breaking of electron-hole symmetry, the diagram is now
a-symmetric around half occupation
of the band. Roughly speaking $s^*$-pairing is favoured far away
from half filling of the band, whereas $d$-wave pairing becomes the
most stable solution near half filling.  We also notice from this plot, that
the a-symmetry implies that the highest $T_c$'s and $d$-pairing
superconductor are to be expected on the left-hand ('hole-doped')
side of half-filling. Lower $T_c$'s and $s$-pairing occur on the right-hand
side.
\\
Let us now consider the $\Delta/T_c$-ratio following from the gap equation.
Within the context of BCS theory we have $\Delta_0(T)=0$ at $T_c$,
so that $T_c$ follows from
 \begin{equation}
 \frac{-2}{U_1}=\sum_q \epsilon_q^{-1}\tanh(\frac{\epsilon_q}{2k_BT_c})
       {[\cos{q_xa} \pm \cos{q_ya}]}^2
 \end{equation}
where the $\pm$ sign refers again to the two symmetries of pairing. This
equation can be easily solved numerically. The result is, that for extended
s-wave pairing the ratio $2\Delta_0/k_BT_c$ is 6.5, whereas for
d-wave pairing it rises gradually from 4 if $|U_1| \ll W$,
up to 6.5 in the limit where $|U_1| \gg W$. This is not sensitive to the
value of the parameter $t'$. We should keep in mind here, that
$\Delta_0$ is the maximum value reached by $\Delta(k)$
(respectively at the $(\pi,0)$- and $(\pi,\pi)$-point for $d$- and
$s^*$-pairing). \\
Let us now look how the mean field estimate of $T_c$ depends on the
coupling strength $|U_1|/W$. In Fig. \ref{fig:tcj0} $T_c^{MF}/W$ is
displayed as a function of $|U_1|/W$ for the d-wave channel. First
of all we notice, that for $|U_1| > W/4$ the value of $T_c^{MF}$ is
about $|U_1|/4$. For $|U_1|/W << 1$ this crosses over to a quadratic dependency
$T_c^{MF}=4|U_1|^2/W$. For
comparison a similar curve is displayed for conventional s-wave pairing,
using the negative U Hubbard model in a band with a square DOS. We notice that
the mean field transition temperature with the latter model
becomes $T_c^{MF}=|U_0|/4$ for large $|U_0|$ (which is actually outside the
range of validity of the BCS weak coupling approach \cite{micnas2,localized}),
and has the familiar BCS-like $\exp{(-W/|U_0|)}$ behaviour for small $U_0$.
The $T_c$ for the extended s-wave pairing lies again below the negative
$U_0$ curve, and is only finite above a threshold value of $|U_1|$ as discussed
above.\\
Let us finally turn to the collective mode spectrum. We can anticipate, that
again d-wave phase fluctuations exist below the particle-hole continuum.
In addition, because there is an on-site repulsive $U_0$, a branch of
spin-fluctuations can be pulled below the particle-hole continuum. In Fig.
\ref{fig:allmodes} the collective mode spectrum is displayed, using
$U_{1}/(4t)=-0.5$, and $n_e=0.85$, and with $U_0/(4t)$ ranging from $0$ to
$1.5$. In the plot for $U_0=0$ we already notice, that the particle-hole
continuum has 8 points in $k$-space where it touches the horizontal axis:
The Fermi surface crosses
the node-lines $k_x=\pm k_y$ at the coordinates
$(\pm (\pi-\delta)/2, \pm (\pi-\delta)/2)$, hence
the particle hole spectrum is gapless for the $Q$-vectors
$(0,\pm (\pi-\delta))$,
$(\pm (\pi-\delta),0)$ and $(\pm (\pi-\delta), \pm (\pi-\delta))$.
Precisely for these
$Q$-values the spin (and charge) susceptibility acquires the largest value,
also in the
superconducting state, hence if we switch on a finite value of the repulsive
on-site $U_0$, a spin-density wave starts to develope around
the $(\pm \pi, \pm \pi)$ points on the Fermi surface.
Clearly the ground-state
is no longer of the form of Eq. \ref{eq:psi},
and the corrections may become strong
enough to completely destroy superconductivity.
As, on the other hand, the spin-density wave exists around a portion
of the Fermi surface where the gap is zero (and therefore contributes the
least to the ground-state energy),  whereas the maximum gap-value is
at the $[\pm \pi,0]$ and $[0, \pm \pi]$ points, there may actually be
a coexistance of superconductivity and a spin-density wave in different
portions of the Fermi surface. \\
{}From Fig. \ref{fig:allmodes} we can see, that the region taking part in the
formation of the spin-density wave quickly spreads around the
$(\pm (\pi-\delta)/2, \pm (\pi-\delta)/2)$ points if $U_0/(4t)$ increases,
leaving a small
region around $[\pm \pi,0]$ and $[0, \pm \pi]$ for the formation of a
superconducting condensate if $U_0/(4t)=1$. The phasediagram for
$U_0/(4t)=1$, and $U_1/(4t)=-0.5$ is indicated in Fig. \ref{fig:phase2}b. The
shaded area roughly indicates the region with an instability towards a
SDW. In principle a mixed
SDW-superconducting state may exist for all concentrations.
It is not possible to decide from the numerical results presented above
whether or not there is a sharp phase boundary separating regions
with a magnetic instability from superconducting regions.
\section{Conclusions}
A unified approach is presented to the calculation of the collective
modes of spin, charge, phase and amplitude in superconductors with
a non-trivial pairing interaction. The expressions for the
dynamical spin- and charge-susceptibilities are generalized to take
into account superconductivity at general values of momentum and frequency.
Several examples are treated. Notably the response functions of a layered
charged electron gas, with a pairing interaction in the $d$-wave channel
is considered in the absence and presence of an on-site Hubbard repulsive
interaction. An incipient instability toward a spin-density wave
follows from the softening of the collective mode spectrum near
$\vec{Q}=(\pi,\pi)$ in the d-wave paired state.
\section{Acknowledgements}
This investigation was supported by the Netherlands Foundation for
Fundamental Research on Matter (FOM) with financial aid from
the Nederlandse Organisatie voor Wetenschappelijk Onderzoek (NWO).

\figure{Diagrams taken into account in the RPA. Exchange self energy (a),
particle-hole mixing (a'), polarization vertex (b and b'), exchange
scattering (c and c'), and direct particle-particle scattering (d and d').
Diagrams b', c' and d' exist only in the superconducting state.
\label{fig:diagrams}}

\figure{Collective mode spectrum of a superconducting layered electron
gas, assuming $s$-wave pairing. The parameters are:
$E_F/(4t)=0.35$, $Ry$/(4t) = 4.0. $Q_{\perp}c$ is varied with 0 to $\pi$
with increments of $0.2 \pi$ (top to bottom solid curves). The dashed
curves are the boundaries of the region of Landau damping. (a) normal
metal, $U$=0 and (b) superconducting state, $U/(4t)=-0.67$.
(c) The amount of $\rho$ (solid) and $\phi$ (dashed) character of the
collective modes in Fig. \ref{fig:layerdispersion} as a function of
collective mode energy. The interruption occur where the modes become
Landau damped.
\label{fig:layerdispersion}}

\figure{Phase diagram in the $U_1$-$n$ plane, where $n$ is the number of
electrons per unit cell, with $t'=-0.7$ and $U_0/(4t)=0$.
\label{fig:phase1}}

\figure{Phase-fluctuating collective mode
versus momentum for a layered electron gas with long range Coulomb
interactions ($Ry^*=20$), an on-site repulsive interaction ($U_0/W=0.5$),
and a nearest neighbour attractive interaction $U_1/W=-0.5$.
The number of electrons is $n_e=0.2$ (a), $n_e=0.25$ (b), $n_e=0.3$ (c)
and $n_e=0.4$ (d).
\label{fig:softphason}}

\figure{(a) Phase diagram in the $T$-$n$ plane, where $n$ is
the number of electrons per
 unit cell, with $t'=-0.7t$ and $U_0/(4t)=0$, and $U_1/(4t)=-0.5$.
 (b) The same
 with $U_0/(4t)=1$
\label{fig:phase2}}

\figure{Solid curve: $T_c/|U_1|$ calculated for the d-wave channel
of the exchange-only model
 with $t'=0$ and 1 electron per site. The same curve is obtained
 for $t'=-0.7t$ with 0.7 electron per site. Open
 lozenges: $T_c$ of the $s^*$-wave channel with the latter parameters.
 Dotted curve:
 $T_c/|U|$ versus $|U|/W$ for the negative $U$ Hubbard model
 taking a square DOS.
\label{fig:tcj0}}

\figure{The collective modes in the d-wave paired state, using
$U_{1}/(4t)=-0.5$, and $n_e=0.85$, and with $U_0/(4t)=0$ (a), $0.5$ (b),
$1$ (c), and $1.5$ (d).
\label{fig:allmodes}}

\end{document}